\documentclass[conference,a4paper]{IEEEtran}
\usepackage[utf8]{inputenc}
\usepackage[normalem]{ulem}
\usepackage{amsmath}
\usepackage{amsfonts}
\usepackage{amssymb}
\usepackage{algpseudocode}
\usepackage{algorithm}
\usepackage[pdftex]{graphicx}
\usepackage{epstopdf,epsfig}
\usepackage{enumerate}
\usepackage{pgfplots}
\usepackage{bbold}
\usepackage{subfigure}
\usepackage{mathtools}
\usepackage{amsthm}
\usepackage{wrapfig}
\usepackage{capt-of}
\usepackage[english]{babel}
\usepackage{url}
\usepackage{color}

\ifCLASSOPTIONcompsoc
  \usepackage[nocompress]{cite}
\else
  \usepackage{cite}
\fi

\theoremstyle{remark}

\theoremstyle{theorem}

\newtheorem{theorem}{Theorem}[section]
\newtheorem{lemma}[theorem]{Lemma}

\theoremstyle{definition}
\newtheorem{definition}{Definition}[section]

\newcommand{\Set}{\mathcal}
\title{Preserving Privacy while Broadcasting: $k$-Limited-Access Schemes}
\author{
\IEEEauthorblockN{Mohammed Karmoose, Linqi Song, Martina Cardone, Christina Fragouli}
University of California Los Angeles, Los Angeles, CA 90095 USA\\
Email: \{mkarmoose, songlinqi, martina.cardone, christina.fragouli\}@ucla.edu}

\begin{document}

\maketitle
\begin{abstract}
Index coding employs coding across clients within the same broadcast domain. 
This typically assumes that all clients 
learn the coding matrix so that they can decode and retrieve their requested data. 
However, learning the coding matrix can pose privacy concerns: it may enable clients to infer information about the requests and side information of other clients~\cite{karmoose2017private}. 
In this paper, we formalize the intuition that the achieved privacy can increase by decreasing the number of rows of the coding matrix that a client learns. 
Based on this, we propose the use of $k$-limited-access schemes: given an index coding scheme that employs $T$ transmissions, we create a $k$-limited-access scheme with $T_k\geq T$ transmissions, and with the property that each client learns at most $k$ rows of the coding matrix to decode its message. We derive upper and lower bounds on $T_k$ for all values of $k$, and develop deterministic designs for these schemes 
for which $T_k$ has an order-optimal exponent for some regimes.
\end{abstract}

\section{Introduction}

Consider a server broadcasting publicly available messages to clients, for instance YouTube videos.
It is well recognized that the use of coding and side information can offer significant bandwidth savings when broadcasting~\cite{bar2011index}. 
However, it can also pose privacy concerns~\cite{karmoose2017private}: although messages are publicly available, clients may wish to preserve the anonymity of their requests from other clients.
A curious client, by leveraging the broadcast information, may be able to infer what
the requests and side information
of other clients are. 
In this paper, we propose new schemes that seek to balance the bandwidth benefits that coding offers with privacy considerations.

We pose this problem within the index coding framework~\cite{bar2011index}.
In index coding, a server has $m$ messages and can losslessly broadcast to $n$ clients. Each client requests a specific message and may have a subset of the messages as side information. To satisfy all clients with the minimum number of transmissions $T$, the server can send coded broadcast transmissions; the clients then use the coding matrix\footnote{The coding matrix has size $T\times m$ and collects in each row the coding coefficients used for the corresponding broadcast transmission.} to decode their messages. In~\cite{karmoose2017private}, we showed that, by knowing the coding matrix, a curious client can  infer information about the side information and requests of other clients.

This paper builds on a new observation: it may not be necessary to provide clients with the entire coding matrix, but with only the rows required for them to decode their own message. 
For example, assume we have $m=4$ messages and $n=4$ clients, where client $i \in \{1,3\}$ has message $b_i$ and would like to receive message $b_{i+1}$, and client $i \in \{2,4\}$ has message $b_i$ and would like to receive message $b_{i-1}$.  The server can satisfy all clients with two broadcast coded transmissions, namely $b_1+b_2$ and $b_3+b_4$, i.e., it uses a $2\times 4$ coding matrix. 
To decode their message, 
 clients $1$ and $2$ only need to know the first row of this matrix (the fact that the first combination is $b_1+b_2$), and similarly clients $3$ and $4$ only need to know the second row of the matrix. 
By restricting the access to the coding matrix, we limit the privacy leakage: the less rows a client learns, the less it can infer about other clients. 

We turn around this observation and ask: what if we restrict each user to access at most $k$ rows of the coding matrix? In particular, assume we are given a coding matrix that uses $T$ transmissions to satisfy all clients.
 Can we ``transform'' it into an ``equivalent'' coding matrix that potentially uses $T_k\ge T$ transmissions to satisfy all clients, but where now each client needs to learn at most $k$ rows of it to decode its message? 
We refer to the coding schemes that satisfy this condition as $k$-limited-access schemes and we evaluate their benefits, cost and feasibility.
 Our main contributions are:
 \begin{enumerate}
\item {\it Benefits:} 
we formalize the intuition that the achieved level of privacy can increase by decreasing the number of rows of the coding matrix that a client learns.
\item {\it Cost:} we derive upper and lower bounds on $T_k$ that highlight the maximum and minimum cost to pay in terms of additional broadcast transmissions as a function of $k$.
 \item {\it Feasibility:} we propose deterministic designs for $k$-limited-access schemes, for all values of $k$. 
For some regimes, our designs provide values of $T_k$ whose exponents are order-optimal.
\end{enumerate}

The paper is organized as follows.
Section~\ref{sec::setup} defines the problem setup.
Section~\ref{sec:MainRes} presents our main results, i.e., it formalizes the intuition that
privacy benefits can be achieved by limiting clients' access to the coding matrix, and
it provides upper and lower bounds on the number of transmissions needed to satisfy clients when they know only part of the matrix.
Section~\ref{eq:ConstrUB} proves the upper bounds presented in Section~\ref{sec:MainRes} by designing $k$-limited-access schemes and assessing their performance. Section~\ref{sec:related} positions our work with respect to related literature and Section~\ref{sec::conclusion} concludes the paper.

\section{Setup and Problem Formulation}
\label{sec::setup}

\smallskip
\noindent\textbf{Notation.} 
Calligraphic letters indicate sets;
boldface lower case letters denote vectors and boldface upper case letters indicate matrices;
$|\Set{X}|$ is the cardinality of $\Set{X}$;
$[n]$ is the set of integers $\{1,\cdots,n\}$;
for all $x \in \mathbb{R}$, the floor and ceiling functions are denoted with $\lfloor x \rfloor$ and $\lceil x \rceil$, respectively;
$\mathbf{0}_{j}$ is the all-zero row vector of dimension $j$;
$\mathbf{1}_{j}$ denotes a row vector of dimension $j$ of all ones and $\mathbf{I}_{j}$ is the identity matrix of dimension $j$;
$\mathbf{e}^j_i$ is the all-zero row vector of length $j$ with a $1$ in position $i$;
logarithms are in base 2.

\smallskip
\noindent\textbf{Index Coding.} We consider an index coding instance, where a server has a database $\mathcal{B}$ of $m$ messages $\mathcal{B}=\left \{ \mathbf{b}_{\Set{M}} \right \}$, where $\Set{M} = [m]$ is the set of message indices, and $\mathbf{b}_j \in \mathbb{F}_{2}^{F}, j \in \mathcal{M},$ with $F$ being the message size. 
The server is connected through a broadcast channel to a set of clients $\mathcal{C}=\left \{ c_{\Set{N}} \right \}$, where $\Set{N} = [n]$ is the set of client indices. 
We assume that $m \geq n$.
Each client {$c_i, i \in \mathcal{N},$} has a subset of the messages {$\left \{ \mathbf{b}_{\Set{S}_i}\right \}$, with $\Set{S}_i \subset \Set{M}$,} as side information and requests a new message {$\mathbf{b}_{q_i}$} with $q_i \in \Set{M} \setminus \Set{S}_i$ that {it} does not have. 
We assume that the server employs a \textit{linear code}, i.e., it designs a set of broadcast transmissions that are linear combinations of the messages in {$\mathcal{B}$}.  The linear index code can be represented as 
{$\mathbf{A} \mathbf{B} = \mathbf{Y}$,}
where {$\mathbf{A} \in \mathbb{F}_2^{T \times m}$} is the coding matrix, $\mathbf{B} \in \mathbb{F}_2^{m \times F}$ is the matrix of all the messages and $\mathbf{Y} \in \mathbb{F}_2^{T \times F}$ is the {resulting matrix} of linear combinations.
Upon receiving {$\mathbf{Y}$,} client $c_i{, i \in \mathcal{N},}$ employs linear decoding
 to retrieve {$\mathbf{b}_{q_i}$.}

\smallskip

\noindent\textbf{Problem Formulation.}
In~\cite{bar2011index}, it was shown that the index coding problem is equivalent to the rank minimization of an $n \times m$ matrix  $\mathbf{G} \in \mathbb{F}_2^{n \times m}$ whose $i$-th row $\mathbf{g}_i$, $i \in [n],$ has the following properties: (i) has a $1$ in the position $q_i$, (ii) has a $0$ in the $j$-th position for all $j \in  \Set{M} \setminus \Set{S}_i$, (iii) can have either $0$ or $1$ in all the remaining positions.
With
this representation, $c_i$ can successfully decode $\mathbf{b}_{q_i}$ using a linear combination of the messages
 corresponding to the non-zero entries of $\mathbf{g}_i$.
Finding an optimal linear coding scheme (i.e., with minimum number of transmissions) is equivalent to completing $\mathbf{G}$ so that it has the minimum possible rank. 
Once we have one such $\mathbf{G}$, we can use a basis of the row space of $\mathbf{G}$ (of size $T = \text{rank} \left (\mathbf{G} \right)$) as coding matrix  $\mathbf{A}$. 
In this case, in fact, client $c_i$ can construct $\mathbf{g}_i$ as a linear combination of the rows of $\mathbf{A}$, i.e., $c_i$ performs the decoding operation $\mathbf{d}_i \mathbf{A} \mathbf{B} = \mathbf{d}_i \mathbf{Y}$, where $\mathbf{d}_i \in \mathbb{F}_2^T$ is the decoding row vector of $c_i$ chosen such that $\mathbf{d}_i \mathbf{A} = \mathbf{g}_i$.
We remark that any index coding scheme that satisfies all clients with $T$ transmissions (where $T$ is not necessarily optimal) -- and can be obtained by any index code design algorithm~\cite{esfahanizadeh2014matrix,huang2015index,chaudhry2008efficient} -- 
corresponds to a completion of $\mathbf{G}$
(i.e., given $\mathbf{A} \in \mathbb{F}_2^{T \times m}$, we can create a corresponding $\mathbf{G}$ in polynomial time).

In our problem formulation we assume we start with a given matrix $\mathbf{G}$ of rank $T$, i.e.,
we are given $n$ {\it distinct} vectors that belong to a $T$-dimensional subspace.
Using a basis of the row space of the given $\mathbf{G}$, we construct $\mathbf{A} \in \mathbb{F}_2^{T\times m}$.
Then, we ask:\\
\textit{Given $n$ distinct vectors  $\mathbf{g}_i$, $i=[n]$, in a $T$-dimensional space, can we find a minimum-size set $\mathcal{A}_k$ with $T_k\geq T$ vectors,
such that each $\mathbf{g}_i$ can be expressed as a linear combination of at most $k$ vectors in $\mathcal{A}_k$ (with $1 \leq k \leq T$)?}\\
The vectors in  $\mathcal{A}_k$ form the rows of the coding matrix $\mathbf{A}_{k}$ we will employ.
We can equivalently restate this as follows.\\
\textit{Given a coding matrix  $\mathbf{A}$,
can we find $\mathbf{P} \in \mathbb{F}_2^{T_k \times T}$, with $T_k$ as small as possible, such that
${\mathbf{A}_{k}} = \mathbf{P} \mathbf{A}$
and each row in $\mathbf{G}$ can be reconstructed by combining at most $k$ rows of $\mathbf{A}_{k}$?}\\                                                               
Note that $k = T$ corresponds to the conventional transmission scheme of an index coding problem for which $\mathbf{P} = {\mathbf{I}_{T}}$.

\smallskip

\noindent\textbf{Transmission Overhead.}
We note that the server can privately share the (at most) $k$ coding vectors that each $c_i$ needs by using a private secret key or a dedicated channel (e.g., the same channel used by $c_i$ to convey the request $q_i$ to the server).  
Thus, using a $k$-limited-access scheme  incurs an extra transmission overhead to privately convey the coding vectors.
In particular, the total number of transmitted bits $\text{C}_k$  is upper bounded by
$\text{C}_k \leq n k m + T_k F,$
while the total number of transmitted bits 
$\text{C}$ using a conventional scheme is $\text{C} = T (F+m)$. 
We observe that the extra overhead incurred  is negligible in comparison to the broadcast transmissions that convey the encoded messages when $n$ and $m$ are both $o(F)$, which is a reasonable assumption for large file sizes (for instance, when sharing YouTube videos).

\section{Main Results}
\label{sec:MainRes}

Consider the setup in the previous section and suppose that client $c_1$ is curious, i.e., by leveraging the $k$ (linearly independent) rows of $\mathbf{A}_k$ that it receives, it seeks to infer information about $c_i, i \in [n], i \neq 1$.
We are interested in quantifying the amount of information that $c_1$ can obtain about $q_i$ (i.e., the identity of the request of $c_i$) as a function of $k$.

As a first step towards this end, we define our privacy metric as follows.
We assume that the index coding instance is random and we 
let $L$ (respectively, $L_1$) be the random variable associated with the subspace spanned by the $T$ rows of the  coding matrix $\mathbf{A} \in \mathbb{F}_2^{T\times m}$ (respectively, spanned by the $k$ vectors given to $c_1$).
Assume that $c_1$ knows $T$.
Then,
\begin{definition}
The privacy metric is defined as $H \left(L|L_1, T\right)$, i.e., it quantifies the amount of uncertainty (entropy) that $c_1$ has about the subspace spanned by the $T$ rows of the index coding matrix $\mathbf{A}$.
\end{definition}
The main motivation behind our choice of the privacy metric is that it offers a yardstick for evaluating the amount of information that $c_1$ can obtain about $q_i$.
This is because $\mathbf{g}_i \in \mathbb{F}_2^m$ (that $c_i$ needs to recover $\mathbf{b}_{q_i}$) lies in the subspace spanned by the $T$ rows of $\mathbf{A}$.
Then, given the specific realizations $T=t$ and $L_1 = \ell_1$, we compute
\begin{align}
P_k &= H \left( L | L_1 = \ell_1, T=t  \right ) 
\stackrel{{\rm{(a)}}}{=} \log \left( |\mathcal{L} (t, \ell_1)|\right) \nonumber
\\ &\stackrel{{\rm{(b)}}}{=}\log \left(  \prod_{\ell=0}^{t-k-1} \frac{2^m-2^{k+\ell}}{2^t-2^{k+\ell}} \right )
\stackrel{m \gg t}{\approx} m(t-k),
\label{eq:ent}
\end{align}
where:
(i) in $\rm{(a)}$ we let $\mathcal{L}(t, \ell_1)$ represent the set of subspaces $L_t \subset \mathbb{F}_2^m$ of dimension $t$ such that $\ell_1 \subset L_t$;
moreover, the equality follows by assuming that the underlying system maintains a uniform distribution across all feasible $t$-dimensional subspaces of $\mathbb{F}_2^m$;
(ii) the equality in ${\rm{(b)}}$ follows by standard counting arguments used to characterize the number of distinct subspaces of a given dimension in a vector space. 
It is clear that, when $m \gg t$, then $P_k$ in~\eqref{eq:ent} decreases linearly with $k$, i.e., the less rows of the coding matrix $c_1$ learns, the less it can infer about the subspace spanned by the $T$ rows of the coding matrix $\mathbf{A}$.
This suggests that, by increasing $k$, $c_1$ has more uncertainty about $q_i$.
It is also clear that $P_k$ in~\eqref{eq:ent} is zero when $k=t$; this is because, under this condition, $c_1$ receives the entire index coding matrix and hence it will be able to perfectly reconstruct the subspace spanned by its rows.
However, although $P_k$ in~\eqref{eq:ent} is zero when $k=t$, $c_1$ might still have uncertainty about $q_i$~\cite{karmoose2017private}.
Quantifying this uncertainty is an interesting open problem that does not appear to be an easy task; this uncertainty, in fact, depends on the underlying system, e.g., on the index code used by the server and on the distribution with which the index coding matrix is selected.

We now build on the analysis above -- that shows the benefits of limiting the access of the clients to the coding matrix -- and focus on finding conditions that guarantee that $\mathbf{P}$ can be constructed while ensuring that each client $c_i, i \in [n],$ successfully decodes its request $\mathbf{b}_{q_i}$ using at most $k$ transmissions.
Towards this end, we derive upper and lower bounds on $T_k$. 
In particular, our main result is stated in the theorem below.
\begin{theorem}
\label{theorem_main}
Given an index coding matrix $\mathbf{A} \in \mathbb{F}_2^{T \times m}$ with $T \geq 2$, it is possible to transform it into ${\mathbf{A}_{k}} = \mathbf{P} \mathbf{A}$ with $\mathbf{P} \in \mathbb{F}_2^{T_k \times T}$, such that each client can recover its request by combining at most $k$ rows of it, if and only if
\begin{align}
 T_k \geq T^\star= \min \left\{T_k : \sum\limits_{i=1}^{k} {T_k \choose i} \geq n \right\}. \label{eq::lb}
\end{align}
Moreover, there exist constructions of $\mathbf{P}$ such that:
\begin{itemize}
\item When $ \lceil T/2 \rceil \leq k < T$, then 
\begin{align}
T_k  \leq \min \left\{n, T + 1 \right\};  \label{theorem_ub2}
  \end{align}
\item When $ 1 \leq k < \lceil T/2 \rceil$, then
\begin{align}
 T_k \leq \min \left\{n,T_{\text{ub}} \right\},  \label{theorem_ub1}
\end{align}
where 
 \begin{align}
  &T_{\text{ub}} \!=\!  \left\{ \begin{array}{ll} 2^T & \text{if } k \!=\! 1 \\
                      2(2k+1)^{\left\lceil \frac{T}{2k-1} \right\rceil - 1} & \text{if } T_{\text{last}} \!=\! 1 \\
                      (2k+1)^{\left\lceil \frac{T}{2k-1} \right\rceil - 1} \left(T_{\text{last}} \!+\! 2  \right) & \text{otherwise}
                     \end{array} \right. \label{T_ub} \\
  & \quad = 2^{O\left({\frac{T}{k}\log k }\right)} \: \text{ if } k \neq 1, \nonumber
 \end{align}
where $T_{\text{last}} = T - (2k-1) \left(\left\lceil \frac{T}{2k-1} \right\rceil -1\right)$.
\end{itemize}
 \end{theorem}

{We provide the proof of the lower bound in~\eqref{eq::lb}
in the Appendix,} while in Section~\ref{eq:ConstrUB} we {give} explicit constructions for $\mathbf{P}$ for the two regimes in Theorem~\ref{theorem_main}, hence proving the upper bounds on $T_k$ in~\eqref{theorem_ub2} and~\eqref{theorem_ub1}.
The results in Theorem~\ref{theorem_main} also imply the following lemma (see also the Appendix).

\begin{lemma}
\label{lemma:SpecCase}
Consider the regime $n = 2^T - 1$. We have
\begin{itemize}
\item When $ \lceil T/2 \rceil \leq k < T$, the bounds in~\eqref{eq::lb} and~\eqref{theorem_ub2} coincide, i.e., the provided construction of $\mathbf{P}$ is optimal;
\item When $ 1 \leq k < \lceil T/2 \rceil$, then the bound in~\eqref{eq::lb} becomes
\begin{align}
  T_k \geq \dfrac{k}{e} \left( \dfrac{2^T - 1}{k} \right)^{1/k} \!=\! 2^{ \Omega \left (\frac{T}{k} + \alpha \log k \right)}, \: \alpha \!=\! \frac{k-1}{k}. \label{theorem_lb2}
 \end{align}
\end{itemize}
\end{lemma}

We now conclude this section with some comparisons between the lower and upper bounds on $T_k$ for the case $n=2^T-1$.
According to Lemma~\ref{lemma:SpecCase}, a construction of $\mathbf{P}$ with $T_k = T+1$ (provided in Section~\ref{eq:ConstrUB}) is optimal for $k \geq \left\lceil T/2 \right\rceil$. 
In other words, by adding only one more transmission to the original index code, clients need {\it at most} half of the transmissions to recover their request; 
this enhances
the attained level of privacy.
Differently, for $1 \leq k < \lceil T/2 \rceil$, the orders of the lower bound in~\eqref{theorem_lb2} and upper bound in~\eqref{theorem_ub1} are different.
This implies that the construction of $\mathbf{P}$ for this regime (see Section~\ref{eq:ConstrUB} for the details) is not optimal.
However, we show next that there exist some regimes of $k$ where the two bounds are close in order.
In particular,


\noindent\textbf{$\bullet$ $k$ is constant.} 
In this case, we have $T_k = 2^{\Theta(T)}$, i.e., the upper and lower bounds have the same order in the exponent.


\noindent \textbf{$\bullet$ $k = T/c$ where $c>1$ is constant.} 
In this case, we have $T_k = 2^{\Theta(\log T)}$, i.e., this represents another regime where the upper and lower bounds have the same order in the exponent.
%
%


\section{Constructions of $k$-Limited-Access Schemes}
\label{eq:ConstrUB} 

In this section, we give explicit constructions of the $\mathbf{P}$ matrix {and prove the two upper bounds on $T_k$ in~\eqref{theorem_ub2} and~\eqref{theorem_ub1}.} 
Our design of $\mathbf{P}$ allows to reconstruct any of the $2^T$ vectors of size $T$.
Recall that $\mathbf{A}$ is full rank and that the $i$-th row of $\mathbf{G}$ can be expressed as $\mathbf{g}_i = \mathbf{d}_i \mathbf{A}$, where $\mathbf{d}_{i} \in \mathbb{F}_2^T$ is the coefficients {row} vector associated {with $\mathbf{g}_i$.}
 
\smallskip

\noindent\textbf{Case I:}
$ \lceil T/2 \rceil \leq k < T$.
When $n \geq T +1$, let
\begin{align}
\label{eq:PPartI}
 \mathbf{P} = \left[ \begin{matrix} \mathbf{I}_{T} \\ \mathbf{1}_T \end{matrix} \right],
\end{align}
\noindent which results in a matrix $\mathbf{A}_k$ with $T_k = T + 1$, {matching} {the bound} in~\eqref{theorem_ub2}.
%
We now show that each $\mathbf{g}_i = \mathbf{d}_i \mathbf{A}, {i \in [n],}$ can be reconstructed by combining up to $k$ vectors of $\mathbf{A}_k$. Let $w (\mathbf{d}_i)$ be the Hamming weight of $\mathbf{d}_i$. If $w(\mathbf{d}_i) \leq \lceil T/2 \rceil$, then we can reconstruct {$ \mathbf{g}_i$} as $\mathbf{g}_i = [\mathbf{d}_i \:\: 0] \mathbf{A}_k$, which involves adding $w(\mathbf{d}_i) \leq \lceil T/2 \rceil \leq k$ rows of $\mathbf{A}_k$. 
{Differently,} if $w(\mathbf{d}_i) \geq \lceil T/2 \rceil + 1$, then we can reconstruct $\mathbf{g}_i$ as ${\mathbf{g}_i} = [\bar{\mathbf{d}}_i \:\: 1] {\mathbf{A}_{k}}$, where $\bar{\mathbf{d}}_i$ is 
the bitwise complement of $\mathbf{d}_i$. 
In this case, reconstructing $\mathbf{g}_i$ involves adding $T - w(\mathbf{d}_i) + 1 \leq \lfloor T/2 \rfloor \leq k$ rows of $\mathbf{A}_k$. 

When $n < T+1$, then it is sufficient to send $n$ uncoded transmissions, where the $i$-th transmission satisfies $c_i, i \in [n]$. 
In this case $c_i$ has access only to the $i$-th transmission, i.e., $k=1$.
This completes the proof of the upper bound in~\eqref{theorem_ub2}.

\smallskip

\noindent\textbf{Case II:}
$1 \leq k < \lceil T/2 \rceil$. 
First, we consider $n \geq T_{\text{ub}}$, where $T_{\text{ub}}$ is defined in \eqref{T_ub}.
For this, we provide a construction for $\mathbf{P}$ that is based on multiple uses of the construction in Case~I. 
In what follows, we let $T_c = \left\lceil \frac{T}{2k-1} \right\rceil$.
Consider the following sets of {\it distinct} vectors (i.e., by omitting replicated vectors)
\begin{subequations}
\label{eq:SETS}
\begin{align}
&\mathcal{P}_j  = \left \{ \mathbf{0}_{2k - 1}, \mathbf{1}_{2k \!-\! 1}, \mathbf{e}_i^{2k-1}; \forall i \!\in\! [2k\!-\!1] \right \},  j \in \left [ T_c \!-\!1 \right ],
\\
&\mathcal{P}_{T_c} = \left \{ \mathbf{0}_{T_{\text{last}}}, \mathbf{1}_{T_{\text{last}}}, \mathbf{e}_i^{T_{\text{last}}}; \forall i \in [T_{\text{last}}] \right \},
\end{align}
\end{subequations}
%
where $T_{\text{last}} = T - (2k-1) (T_c-1)$.
Then, our construction of $\mathbf{P}$ is based on different concatenations of various elements of the above sets as we explain in what follows.
Let
\begin{align*}
\mathcal{P} = \mathcal{P}_1 \times \mathcal{P}_2 \times \ldots \times \mathcal{P}_{T_c}
\end{align*}
be the Cartesian product of the sets defined in~\eqref{eq:SETS} and $\mathcal{P}(i),i \in [T_k],$ be the $i$-th tuple of $\mathcal{P}$.
Then, the $i$-th row vector $\mathbf{p}_i$ of $\mathbf{P}$ is constructed by concatenating the elements of the tuple $\mathcal{P}(i)$ in their respective order (i.e., the first element of $\mathcal{P}(i)$ is the left-most part of $\mathbf{p}_i$, the second element of $\mathcal{P}(i)$ is the second left-most part of $\mathbf{p}_i$ and so on).
It is not difficult to see that with this construction $\mathbf{p}_i$ has length $T$.
%
%
%
%
Since from~\eqref{eq:SETS} we have that, for $j \in [T_c - 1]$,
\begin{align*}
 |\Set{P}_j| = \left\{\begin{array}{ll}
                       2 & k = 1 \\
                       2k+1 & k > 1
                      \end{array}
 \right.  |\Set{P}_{T_c}| = \left\{\begin{array}{ll}
                       2 & T_{\text{last}} = 1 \\
                       T_{\text{last}}+2 & T_{\text{last}} > 1
                      \end{array} \right. ,
\end{align*}
then we have $\prod_{j=1}^{T_c} |\Set{P}_j |$
possible different ways of concatenating vectors from these sets. This gives the bound in~\eqref{theorem_ub1}.
To illustrate this process consider the following example.

\begin{figure}
  \centering
  \includegraphics[width=0.5\columnwidth]{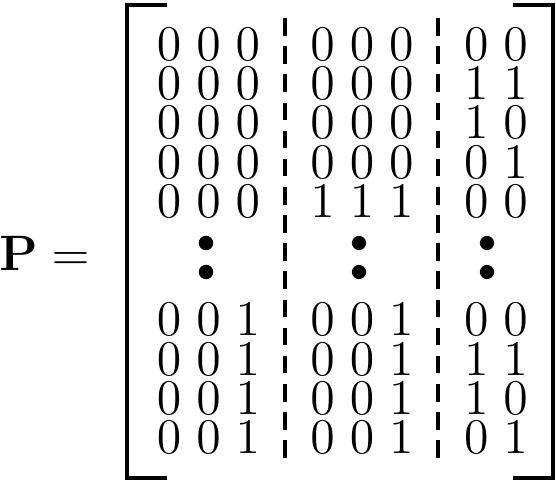}
  \caption{Construction of $\mathbf{P}$ for $T = 8$ and $k = 2$.}
  \label{fig::construct_P_example}
 \end{figure}

\noindent{\it Example.} Let $T = 8$ and $k = 2$ for which $T_c=3$ and $T_{\text{last}}=2$.
Then, we have
\begin{align*}
\mathcal{P}_1 = \mathcal{P}_2 &= \left \{ \begin{bmatrix}0 & 0 &0   \end{bmatrix},\begin{bmatrix}1 & 1 &1   \end{bmatrix}, \begin{bmatrix}1 & 0 &0   \end{bmatrix}, \right.
\\ & \quad \ \; \left.\begin{bmatrix}0 & 1 &0   \end{bmatrix}, \begin{bmatrix}0 & 0 &1   \end{bmatrix} \right \},
\\
\mathcal{P}_3 &= \left \{ \begin{bmatrix}0 & 0    \end{bmatrix},\begin{bmatrix}1 & 1    \end{bmatrix},\begin{bmatrix}1 & 0    \end{bmatrix},\begin{bmatrix}0 & 1    \end{bmatrix} \right \}.
\end{align*}
Figure~\ref{fig::construct_P_example} shows how $\mathbf{P}$ is then constructed. 
%

{
We now need to prove that any $\mathbf{g}_i, i\in [n],$ can be reconstructed using at most $k$ rows of $\mathbf{A}_k$. 
Notice that this is equivalent to showing that we need at most $k$ rows of $\mathbf{P}$ to reconstruct $\mathbf{d}_i, i \in [n]$.
This is because, if this holds, then $\mathbf{d}_i = \mathbf{d}^\star_i \mathbf{P}$ where the row vector $\mathbf{d}^\star_i \in \mathbb{F}_2^{T_k}$ has at most $k$ non-zero elements.
Then, this would imply
$\mathbf{g}_i = \mathbf{d}_i \mathbf{A} = \mathbf{d}^\star_i \mathbf{P} \mathbf{A} = \mathbf{d}^\star_i\mathbf{A}_k$,
i.e., $\mathbf{g}_i$ is reconstructed by using at most $k$ rows of $\mathbf{A}_k$.
In what follows, we therefore prove that any $\mathbf{d}_i, i \in [n],$ can be reconstructed by using at most $k$ rows of $\mathbf{P}$.
As a running example to illustrate the different steps of our proof we use the case in Figure~\ref{fig::construct_P_example} with $\mathbf{d}_i = \begin{bmatrix} 1 & 1 & 0 &  0 & 1 & 0 & 0 & 0\end{bmatrix}$.

\noindent {\bf{Step~1.}}
Starting from the left-most bit, we split $\mathbf{d}_i, i\in[n],$ into $(T_c-1)$ parts of length ($2k-1$) and one last part of length $T_{\text{last}}$.
We denote by $\mathbf{d}_i(j)$ the $j$-th part with $j \in  [T_c]$.

\noindent {\it{Running example.}} We have
\begin{align*}
\mathbf{d}_i(1) = \begin{bmatrix} 1 & 1 & 0\end{bmatrix}, \ \mathbf{d}_i(2) = \begin{bmatrix} 0 & 1 & 0\end{bmatrix}, \
\mathbf{d}_i(3) = \begin{bmatrix} 0 & 0\end{bmatrix}.
\end{align*}

\noindent {\bf{Step~2.}}
We leverage our proof of Case~I, where we showed that $\mathbf{P}$ in~\eqref{eq:PPartI} can be used to reconstruct any vector of length $T$ using $ \lceil T/2 \rceil \leq k < T$ rows.
This, in fact, implies that:
(i) any $\mathbf{d}_i(j), j \in [T_c-1],$ can be reconstructed by adding at most $k$ elements of $\mathcal{P}_j$ (excluding the first element), and
(ii) any $\mathbf{d}_i(T_c)$ can be reconstructed by adding at most $k$ elements of $\mathcal{P}_{T_c}$ (excluding the first element).
We let $\mathcal{R}_j, j \in [T_c],$ be the set of elements of $\mathcal{P}_j$ needed to reconstruct $\mathbf{d}_i(j)$.
Clearly, $\left |\mathcal{R}_j \right | \leq k , j \in [T_c]$.
Let $R^\star = \max_{j \in [T_c]} \left |\mathcal{R}_j \right |$.
Then, we further populate $\mathcal{R}_j, j \in [T_c]$ with $R^\star - \left |\mathcal{R}_j \right |$ zero vectors, so that all $\mathcal{R}_j$ have the same cardinality.

\noindent {\it{Running example.}} We have $R^\star = 2$ and
\begin{align*}
\mathcal{R}_1 &= \left \{\begin{bmatrix}1 & 0 & 0\end{bmatrix},\begin{bmatrix}0 & 1 & 0\end{bmatrix} \right \},
\\
\mathcal{R}_2 &= \left \{\begin{bmatrix}0 & 1 & 0\end{bmatrix},\begin{bmatrix}0 & 0 & 0\end{bmatrix} \right \},
\\
\mathcal{R}_3 &=\left \{\begin{bmatrix}0 & 0 \end{bmatrix},\begin{bmatrix}0 & 0 \end{bmatrix} \right \}.
\end{align*}

\noindent {\bf{Step~3.}}
We concatenate the different elements of $\mathcal{R}_j, j \in [T_c]$. 
In particular, for each $\ell \in [R^\star]$ we create a vector of length $T$ by concatenating the elements in the $\ell$-th position of all $\mathcal{R}_j, j \in [T_c],$ as follows: we put the $\ell$-th element of $\Set{R}_1$ as the left-most part, then we concatenate to it the $\ell$-th element of $\Set{R}_2$ and so on until $\Set{R}_{T_c}$.
Thus, we obtain a set $\mathcal{R}^\star$ of $R^\star$ vectors of length $T$. 
Clearly, from our construction of $\mathbf{P}$, each element of $\mathcal{R}^\star$ is a row of $\mathbf{P}$.
Moreover, from our construction in the previous step of $\mathcal{R}_j, j \in [T_c]$, we have that the sum of the $R^\star$ vectors in $\mathcal{R}_j$ reconstructs $\mathbf{d}_i(j)$.
Hence, it is not difficult to see that the sum of the $R^\star$ elements of $\mathcal{R}^\star$ reconstructs $\mathbf{d}_i$. 

\noindent {\it{Running example.}} We have
\begin{align*}
\mathcal{R}^\star & =  \left\{ \begin{bmatrix} 1 & 0 & 0 & 0 & 1 & 0 & 0 & 0 \end{bmatrix}, \right.
\\ & \left. \quad \ \ \begin{bmatrix} 0 & 1 & 0 & 0 & 0 & 0 & 0 & 0 \end{bmatrix} \right \}.
\end{align*}
By adding the two elements of $\mathcal{R}^\star$ we obtain $\begin{bmatrix} 1 & 1 & 0 & 0 & 1 & 0 & 0 & 0 \end{bmatrix}$, which is precisely the $\mathbf{d}_i$ we wanted to reconstruct.

When $n < T_{\text{ub}}$, then it is sufficient to send $n$ uncoded transmissions, where the $i$-th transmission satisfies $c_i, i \in [n]$. 
In this case $c_i$ has access only to the $i$-th transmission, i.e., $k=1$.
This completes the proof of the upper bound in~\eqref{theorem_ub1}.

}

\section{Related Work} \label{sec:related}
The problem of protecting privacy was initially proposed
to enable the disclosure of databases for public access, while maintaining the anonymity of the users~\cite{aggarwal2008general}.
Similar concerns have been raised in the context of
{\it Private Information Retrieval} (PIR),
{which was introduced in~\cite{chor1998private} and has received a fair amount of attention~\cite{freij2016private,banawan2016capacity}.}
In particular, in PIR the goal is to ensure that no information about clients' requests is revealed to a set of malicious databases when clients are trying to retrieve information from them.
Similarly, the problem of {\it Oblivious Transfer} (OT) was studied~\cite{Brassard1987,mishra2014oblivious} to establish, by means of cryptographic techniques, two-way private connections between the clients and the server.

We were here interested in addressing privacy concerns in broadcast domains.
In particular, we analyzed this problem within the index coding framework, as we recently proposed in~\cite{karmoose2017private}.
This problem differs from secure index coding~\cite{dau2012security}, where the goal is to guarantee that each client does not learn any information about the {\it content} of the messages other than its request. 
Differently, our goal was to limit the information that a client can learn about the {\it identities} of the requests of other clients.
Moreover, our approach here has a significant difference with respect to~\cite{karmoose2017private}.
In fact, while in~\cite{karmoose2017private} our goal was to design the encoding matrix to guarantee a high-level of privacy, here we assumed that an index coding matrix (that satisfies all clients) is given to us and we developed methods to increase its achieved levels of privacy.

The solution that we here proposed to limit the privacy leakage  is based on finding overcomplete bases.
This approach 
is closely related to
compressed sensing and dictionary learning~\cite{chen2016compressed}, where the goal is to learn a dictionary of signals such that other signals can be {\it sparsely} and {\it accurately} represented using atoms from this dictionary.
These problems seek lossy solutions, i.e., signal reconstruction is not necessarily perfect.
This allows a convex optimization formulation of the problem, which can be solved efficiently~\cite{rubinstein2010dictionaries}.
{In contrast,} our problem was concerned with lossless reconstructions, in which case the optimization problem is no longer convex.

\section{Conclusion}
\label{sec::conclusion}
We studied an index coding problem, where clients are eager to learn the identity of the request of other clients.
We proposed the use of $k$-limited-access schemes to mitigate the privacy risks, which provide clients with only part of the coding matrix and still ensure that they can all recover their requested message.
We showed that such approach can achieve higher levels of privacy than conventional schemes (where the entire matrix is broadcast to all users) at the cost of additional number of transmissions.
This analysis sheds light on an inherent tradeoff between bandwidth savings and privacy protection in broadcast domains.
%
Future work would include the derivation of tighter upper and lower bounds on the number of transmissions required by a $k$-limited-access scheme.


\section*{Appendix}

In order to prove the lower bound in~\eqref{eq::lb},
we establish a connection between our problem and a linear-algebraic one, namely the problem of representing vectors in finite vector spaces.
Given a matrix $\mathbf{A}$, denote by {$\mathbb{V}_{\mathbf{A}} \subseteq \mathbb{F}_{2}^T$} the subspace formed by the span of the rows of $\mathbf{A}$. 
It is {clear that} the dimension of {$\mathbb{V}_{\mathbf{A}}$} is at most $T$ (exactly $T$ {if $\mathbf{A}$ is full rank)} and that the $n$ distinct rows of $\mathbf{G}$ lie in {$\mathbb{V}_{\mathbf{A}}$.}
Let $\mathbf{a}_i \in \mathbb{F}_2^{m}, i \in [T_k],$ be the $i$-th row of $\mathbf{A}_k$.
Then this problem is equivalent to the following: \textit{what is a minimum-size set of vectors $\Set{A}_{k} = \{ \mathbf{a}_{[T_k]}\}$ such that any row vector of $\mathbf{G}$ can be represented by a linear combination of at most $k$ vectors of {$\Set{A}_{k}$}?}

A lower bound on $T_k$ can be obtained as follows.
Given $\Set{A}_{k}$, there must exist a linear combination of at most $k$ vectors of {$\Set{A}_{k}$} that is equal to each of the $n$ distinct row vectors of $\mathbf{G}$. 
The number of \textit{distinct} non-zero linear combinations of 
up to $k$ vectors 
is at most equal to $\sum\limits_{j=1}^k {T_k \choose j}$. 
Thus, we have
\begin{align}
\label{eq:LBRep}
  \sum\limits_{i=1}^{k} {T_k \choose i} \geq n,
 \end{align}
which gives precisely the bound in~\eqref{eq::lb}.

We now derive the lower bounds in Lemma~\ref{lemma:SpecCase}, i.e., we
evaluate~\eqref{eq::lb} for $n=2^T-1$.
From~\eqref{eq:LBRep}, we obtain
 \begin{align}
  \sum\limits_{i=1}^{k} {T_k \choose i} \geq 2^{T} - 1. \label{lem_2}
 \end{align}
{Since in general $T_k \geq T$, to prove that $T_k \geq T+1$ for $k<T$, it is sufficient to show that we have a contradiction for $T_k=T$.
Indeed, by setting $T_k=T$, the bound in~\eqref{lem_2} becomes
\begin{align*}
\sum\limits_{i=1}^{k} {T \choose i} \geq 2^{T} - 1 = \sum\limits_{i=1}^{T} {T \choose i},
\end{align*}
which clearly is not possible since $k<T$.
Hence, $T_k \geq T+1$ for all $k<T$.}
However, for $1 \leq k < \left\lceil T/2 \right\rceil$, we {can refine this lower bound as follows}
\begin{align*}
  k \left(\dfrac{T_k e}{k} \right)^k &\geq k {T_k \choose k} \geq \sum\limits_{i=1}^{k} {T_k \choose i} \geq 2^{T} - 1 \label{lem_2} \\
   {\implies} T_k &\geq \dfrac{k}{e} \left( \dfrac{2^T - 1}{k} \right)^{1/k}.
 \end{align*}
This concludes the proof of the lower bounds in Lemma~\ref{lemma:SpecCase}.

\bibliographystyle{IEEEtran}
\bibliography{Bib}
%
%

\end{document}